\documentclass{article} 
\usepackage{epsfig}

\usepackage{a4}
\newcommand{\beq}{\begin{equation}}
\newcommand{\eeq}{\end{equation}}
\newcommand{\beqs}{\begin{eqnarray}}
\newcommand{\eeqs}{\end{eqnarray}}

\begin{document}
\title{On $d=4$  Yang-Mills instantons in a \\
spherically symmetric background}
 \author{{\large {\bf Yves Brihaye}}$^{\dagger}$ and
{\large {\bf Eugen Radu}}$^{\ddagger}$  \\ \\
$^{\dagger}${\small  Physique-Math\'ematique, Universite de
Mons-Hainaut, Mons, Belgium} \\
$^{\ddagger}${\small Department of
Mathematical Physics,  National University of Ireland, Maynooth, Ireland}\\  
 }
\date{}
\newcommand{\dd}{\mbox{d}}
\newcommand{\tr}{\mbox{tr}}
\newcommand{\la}{\lambda}
\newcommand{\ka}{\kappa}
\newcommand{\al}{\alpha}
\newcommand{\ga}{\gamma}
\newcommand{\de}{\delta}
\newcommand{\si}{\sigma}
\newcommand{\bomega}{\mbox{\boldmath $\omega$}}
\newcommand{\bsi}{\mbox{\boldmath $\sigma$}}
\newcommand{\bchi}{\mbox{\boldmath $\chi$}}
\newcommand{\bal}{\mbox{\boldmath $\alpha$}}
\newcommand{\bpsi}{\mbox{\boldmath $\psi$}}
\newcommand{\brho}{\mbox{\boldmath $\varrho$}}
\newcommand{\beps}{\mbox{\boldmath $\varepsilon$}}
\newcommand{\bxi}{\mbox{\boldmath $\xi$}}
\newcommand{\bbeta}{\mbox{\boldmath $\beta$}}
\newcommand{\ee}{\end{equation}}
\newcommand{\eea}{\end{eqnarray}}
\newcommand{\be}{\begin{equation}}
\newcommand{\bea}{\begin{eqnarray}}
\newcommand{\ii}{\mbox{i}}
\newcommand{\e}{\mbox{e}}
\newcommand{\pa}{\partial}
\newcommand{\Om}{\Omega}
\newcommand{\vep}{\varepsilon}
\newcommand{\bfph}{{\bf \phi}}
\newcommand{\lm}{\lambda}
\def\theequation{\arabic{equation}}
\renewcommand{\thefootnote}{\fnsymbol{footnote}}
\newcommand{\re}[1]{(\ref{#1})}
\newcommand{\R}{{\rm I \hspace{-0.52ex} R}}
\newcommand{\N}{{\sf N\hspace*{-1.0ex}\rule{0.15ex}%
{1.3ex}\hspace*{1.0ex}}}
\newcommand{\Q}{{\sf Q\hspace*{-1.1ex}\rule{0.15ex}%
{1.5ex}\hspace*{1.1ex}}}
\newcommand{\C}{{\sf C\hspace*{-0.9ex}\rule{0.15ex}%
{1.3ex}\hspace*{0.9ex}}}
\newcommand{\eins}{1\hspace{-0.56ex}{\rm I}}
\renewcommand{\thefootnote}{\arabic{footnote}}

\maketitle

\begin{abstract}
We present arguments for the existence of
self-dual Yang-Mills instantons for several spherically symmetric 
backgrounds with Euclidean signature.
The time-independent Yang-Mills field has finite action and a vanishing 
energy momentum tensor and does not disturb the geometry.
We conjecture the existence of similar solutions for any
nonextremal $SO(3)$-spherically symmetric background.
\end{abstract}

\medskip
{\bf Introduction--}The recent work \cite{Brihaye:2006nk} noticed the existence
of a new type of instanton solution
extremizing the Yang-Mills (YM) action
\begin{eqnarray}
\label{lag0} 
S= -\frac{1}{16 \pi^2 e^2} \int_{ \mathcal{M}} d^4x \sqrt{g}~{\rm Tr}~\{F_{ab} F^{ab}\} ,   
\end{eqnarray} 
(with $e$ the gauge coupling constant)
for an euclideanized Schwarzschild background metric.

This YM instanton solution was found for a SU(2) nonabelian ansatz
with no dependence on the Euclidean time
\begin{eqnarray}
\label{YMansatz}
A_a dx^a=\frac{1}{2} \{ u(r) \tau_3 d\tau+  
 w(r) \tau_1  d \theta
+\left(\cot \theta \tau_3
+ w(r) \tau_2 \right) \sin \theta d \varphi \}, 
\end{eqnarray}
(where the  
$\tau_i$ are the standard Pauli matrices),
described by two 
functions $w(r)$ and $u(r)$ which we shall refer to as magnetic 
and electric potential, respectively.

The configuration  reported in \cite{Brihaye:2006nk}
differs from the well-known Charap-Duff solution \cite{Charap:1977ww},
since
it satisfies a different set of boundary conditions and has a different
value of the action. 

Moreover, a double self-dual Schwarzschild background is not crucial 
for the existence of this new type of solutions.
In this letter we report the existence of 
 YM self-dual instantons with similar properties
 for other Euclidean backgrounds with the same amount of symmetry as
 the Schwarzschild metric.
 These solutions are contructed by directly solving the self-duality equations
 with a suitable set of boundary conditions.

{\bf The model--}
The variation of the action priciple (\ref{lag0}) 
with respect to the gauge connection $A_a$ leads to  the  Yang-Mills equations 
\begin{eqnarray}
\label{YM-eqs}
\nabla_{a}(F^{ab})-i[A_{a},F^{ab}]=0.
\end{eqnarray}
Here we will consider YM fields satisfying the duality condition
$F_{ab}=\pm \frac{1}{2}\sqrt{g}\epsilon_{abcd}F^{cd},$
in which case the YM energy-momentum tensor vanishes.
Thus these self-dual gauge field configurations 
 will not disturb the geometry,
 while the
YM equations (\ref{YM-eqs}) are satisfied automatically.

We consider a general spherically symmetric metric ansatz
\begin{eqnarray}
\label{metric1}
ds^2=  
\sigma^2(r) N(r) d\tau^2 + \frac{dr^2}{N(r)} 
+ R^2(r)(d \theta^2 + \sin^2 \theta d \varphi^2),
\end{eqnarray}
without fixing the metric gauge.
Here $\theta$ and $\varphi$ are the angular
coordinates with the usual range,  $\tau$ corresponds to the Euclidean time, which 
has a periodicity $\beta$, while
the radial coordinate $r$ varies between some $r_0$ and infinity. 
As $r\to \infty$, the euclideanized (thermal-)Minkowski background
is approached, with $R(r) \to r$, $\sigma(r) \to 1$, $N(r) \to 1-2M/r$, with $M$ a positive constant.
This type of metric usually corresponds to analytical
continuations of Lorentzian black holes and particle-like
solutions.

For this metric choice,
the self-duality equations read 
(see also the early work \cite{Boutaleb-Joutei:1979va})
\begin{eqnarray}
\label{SD-eq}
w'=\mp \frac{wu}{\sigma N},~~~R^2 u'=\pm \sigma(1-w^2),
\end{eqnarray}
the  
expression of the YM action being
\begin{eqnarray}
\label{action-sd}
S=\mp \frac{  \beta}{2 \pi e^2} \big[u(1-w^2)\big]\bigg|_{r_0}^\infty~~ ,
\end{eqnarray} 
which equals  the SU(2) Pontryagin charge
\begin{equation}
\label{pont}
 P_{YM}=\frac{1}{32\pi^2}
 \int_{ \mathcal{M}} d^4x \sqrt{g}~ \epsilon^{abcd} ~{\rm Tr} \{F_{ab}  F_{cd} \}.
\end{equation}
Without any loss of generality, we solve the self-duality equations
 by taking the upper sign in 
(\ref{SD-eq}); the anti-instanton solutions are found by 
reversing the sign of the electric potential.

The  expansion of the nonabelian potentials as $r \to \infty$ is
\footnote{ The Charap-Duff instanton solution \cite{Charap:1977ww}
is found for a Schwazschild background 
$N=1-2M/r$, $\sigma=1$, $R(r)=r$ and a different set of 
boundary conditions. It reads
$u(r)=-M/r^2$, $w(r)=\sqrt{N}$, the action being
 $S=-1/e^2$ in the normalization we use (see also \cite{Tekin:2002mt}
 for a discussion of this solution).}
\begin{eqnarray}
\label{inf}
w(r)=\frac{e^{-\Phi r}}{r^{2M-1}}+\dots,~~~
u(r)= \Phi-\frac{1}{r}+\dots,
\end{eqnarray}
with $\Phi$ an arbitrary positive constant.
Thus one may  define nonabelian electric and magnetic charges \cite{Brihaye:2006nk}
computed $e.g.$ according to
\begin{equation}
\label{def-charge}
\pmatrix{ {Q_E}\cr  {Q_M}\cr}
 = {1\over 4\pi} \int dS_{\mu} \, \sqrt{g} \,
{\rm Tr} \Big\{ 
 \pmatrix{ F^{\mu \tau}\cr \tilde F^{\mu \tau} \cr} \tau_3 
 \Big\}. 
\end{equation}
The instanton solutions discussed in this paper have $Q_M=Q_E=1$.

{\bf Self dual instantons in a ''bolt'' background--}
 Following \cite{Brihaye:2006nk}, we start by considering
asymptotically flat background metrics whose fixed
point set of the Euclidean time
symmetry  is of two dimensions (a "bolt") and 
the range of the radial coordinate is restricted
to $r_h\leq r<\infty$, while
\begin{eqnarray}
\label{m-eh} 
\nonumber 
N(r)=N_1(r-r_h)+O(r-r_h)^2,
~~~
\sigma(r)=\sigma_h+ O(r-r_h),~~R(r)=R_h+ O(r-r_h),
\end{eqnarray}
where $N_1,~\sigma_h,~R_h$ are positive constant, determined
by the Einstein equations.

The YM potentials have the following expansion as $r \to r_h$  
\begin{eqnarray}
\label{eh}
w(r)=w_h+\frac{w_h(w_h^2-1)}{R^2_hN_1}(r-r_h)+O(r-r_h)^2,~~
u(r)=\frac{\sigma_h(1-w_h^2)}{R_h^2}(r-r_h)+O(r-r_h)^2,
\end{eqnarray}
with $0\leq w_h\leq 1$.
From (\ref{action-sd}), we find the action of the instanton solutions
\begin{eqnarray}
\label{action-tot}
S=-\frac{ \beta\Phi  {Q_M}}{2 \pi e^2} .
\end{eqnarray}
The properties of the background metric enters here through the expression of $\beta$.
Note that (\ref{SD-eq}) imply the existence of a maximal 
allowed magnitude of the electric
potential
at infinity
for a given 
$r_h$  
\begin{eqnarray}
\label{cond}
\Phi< \int_{r_h}^\infty dr~\frac{\sigma(r)}{R^2(r)}.
\end{eqnarray}

\newpage
\setlength{\unitlength}{1cm}

\begin{picture}(18,7)
\centering
\put(1.3,-1){\epsfig{file=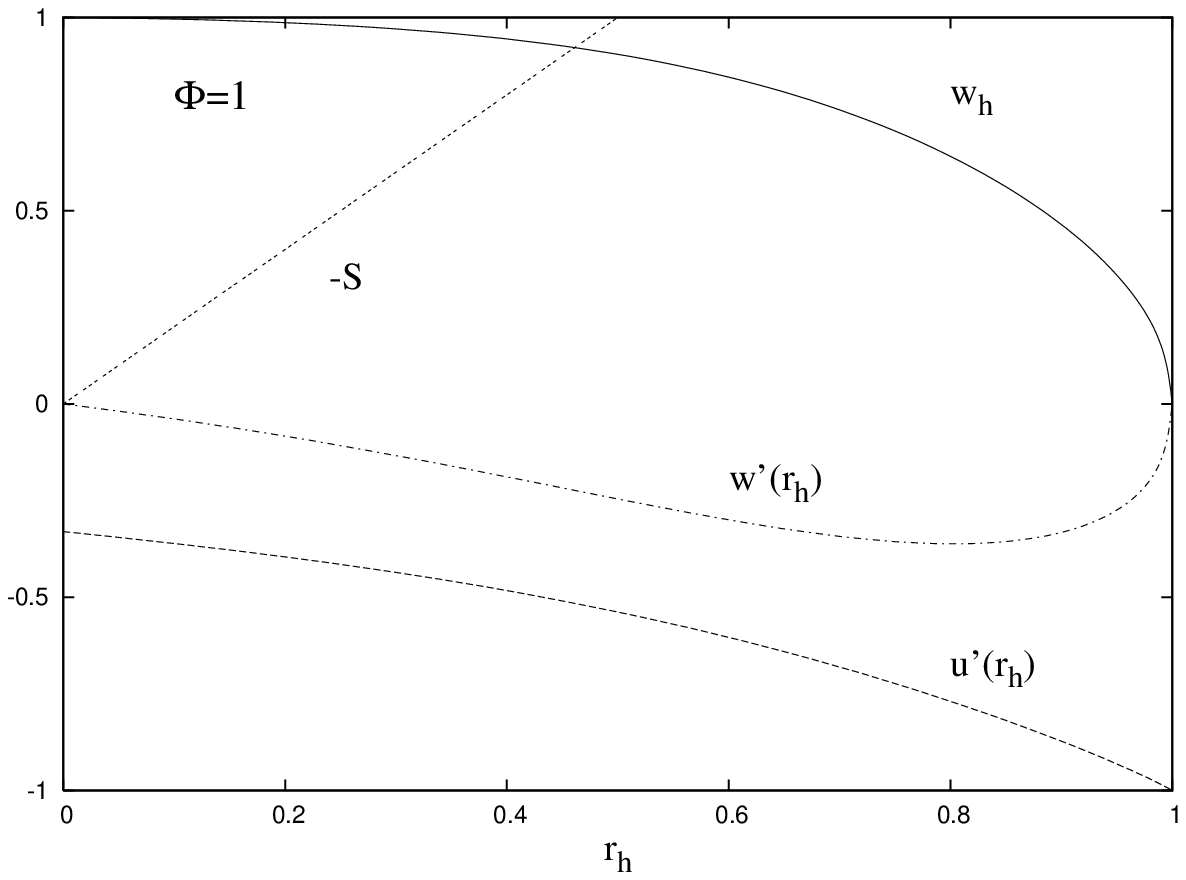,width=11cm}}
\end{picture} 
\\
\\
\\
\\
{\small {\bf Figure 1.} The parameters $w_h,~w'(r_h),~u'(r_h)$ and 
the action $S$ (in units $1/e^2$) of the YM instantons in a Schwarzschild instanton background
are plotted as a function of  $r_h$.}
\\
\setlength{\unitlength}{1cm}

\begin{picture}(18,7)
\centering
\put(1.3,-1){\epsfig{file=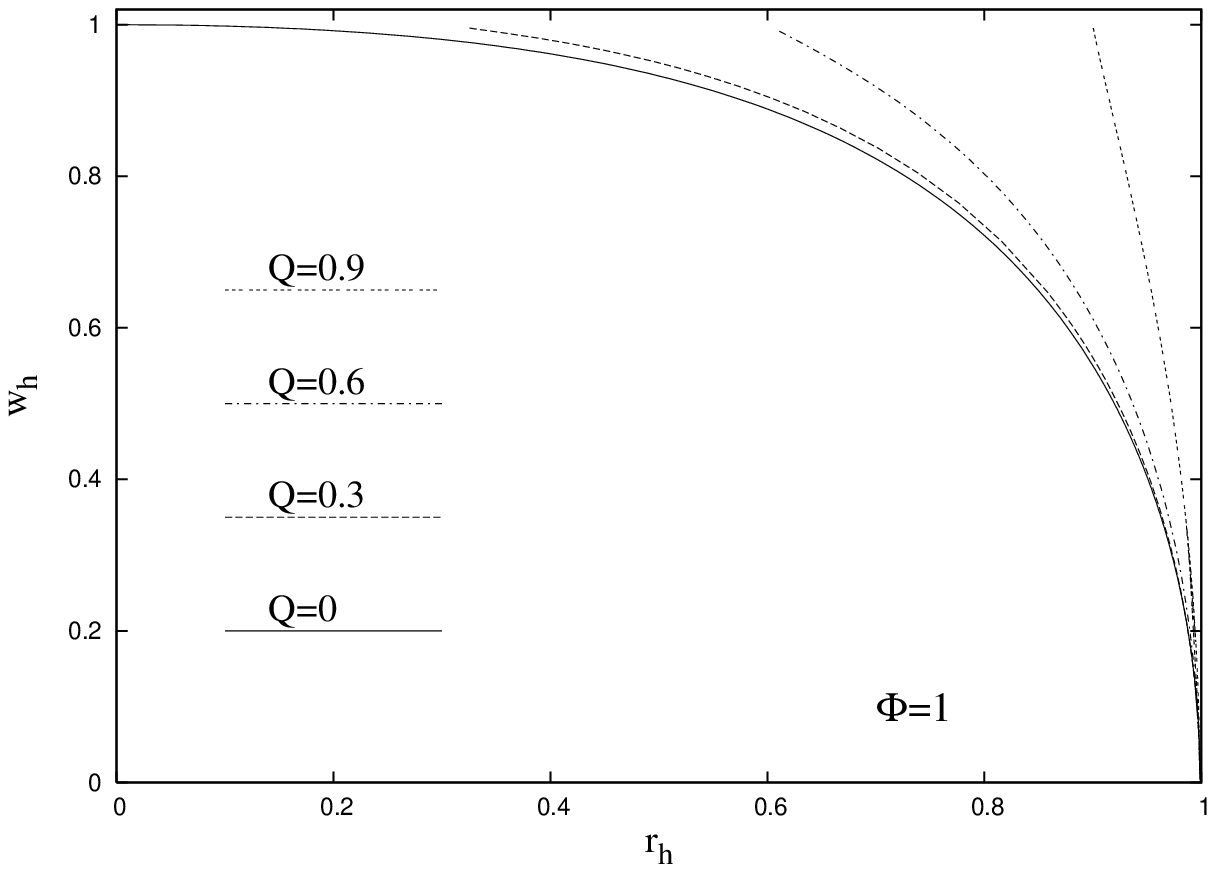,width=11cm}}
\end{picture} 
\\
\\
\\
\\
{\small {\bf Figure 2.} The value $w_h$ of the magnetic
gauge potential at $r=r_h$  is plotted as a function of $r_h$ for 
a Reissner-Nordstr\"om background with different values of the electric charge $Q$.}
\\
\\
\\
In the numerical procedure, we set $\Phi=1$ without
any loss of generality,
which implies $r_h<1$ for both Schwarzschild and Reissner-Nordstr\"om backgrounds.

 Unfortunately, we could not find closed form solutions of the equations
(\ref{SD-eq}) for any physical relevant choice  of the metric backgrounds,
except for the euclideanized Minkowski metric ($i.e.$ $N(r)=\sigma(r)=1$, $R(r)=r$), 
in which case we recover the well-known self-dual YM  solution
 $w=r/\sinh r$, $ u=\coth r-1/r$ \cite{Prasad:1975kr}.
 
\newpage
\setlength{\unitlength}{1cm}

\begin{picture}(18,7)
\centering
\put(1.3,-1){\epsfig{file=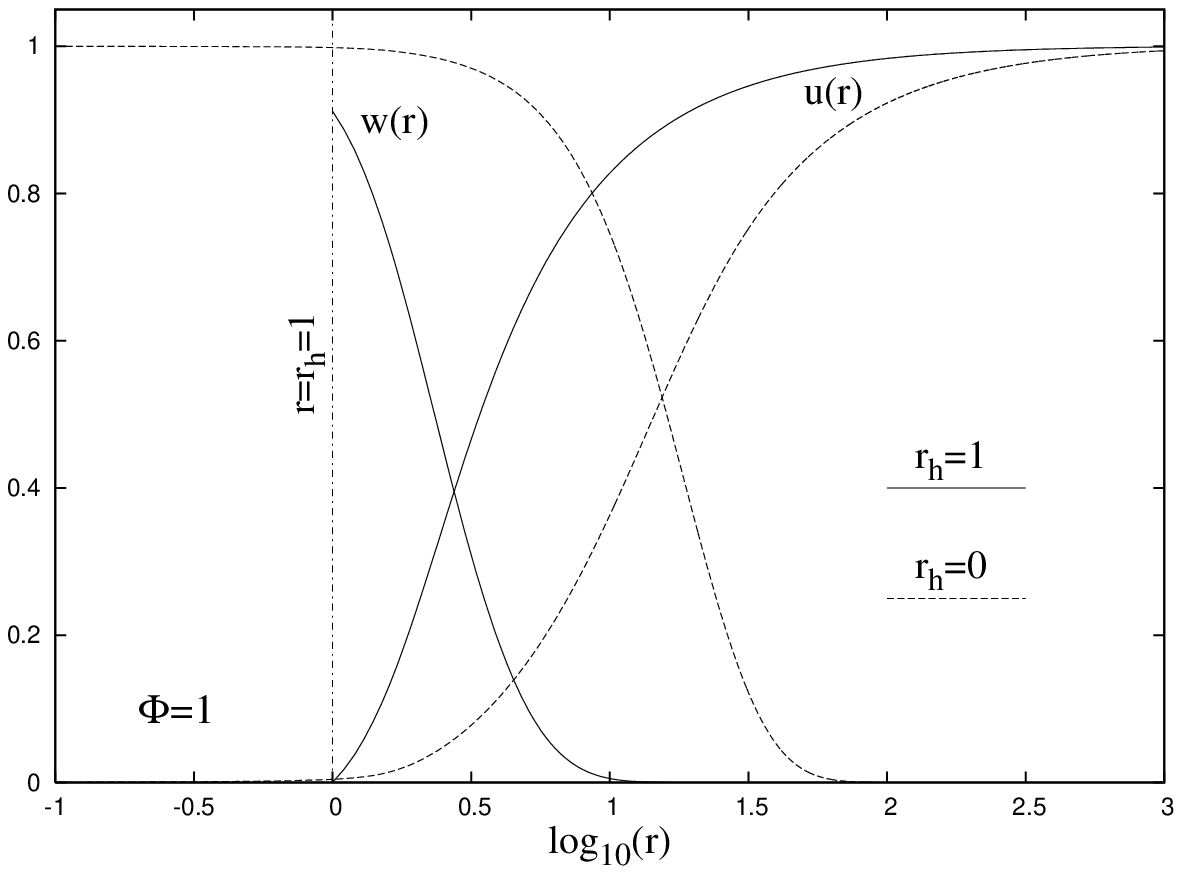,width=11cm}}
\end{picture} 
\\
\\
\\
\\
\\
{\small {\bf Figure 3.} 
The profiles of the gauge functions $w(r)$ and $u(r)$
are plotted for self-dual instantons in the background of   
euclideanized Bartnik-McKinnon solution $(r_h=0)$
and a "bolt" EYM configuration with $r_h=1$. 
The nonself-dual "primary" YM field in the EYM solutions has one node in both cases.
}
\\
\\
However, these equations can easily be solved numerically
\footnote {The numerical integrations were carried out using both a shooting method
as well as applying the numerical program COLSYS~\cite{colsys}, with
complete agreement to very high accuracy.}.

Apart from Schwarzschild instanton, we solved the equations (\ref{SD-eq})
for a Reissner-Nordstr\"om instanton background $(N=1-2M/r+Q^2/r^2,~\sigma=1,
~~R=r)$
\cite{Hawking:1995ap} and
the euclideanized black 
hole solutions of the 
Einstein-Maxwell-dilaton theory \cite{gibbons, horowitz},
with 
\begin{eqnarray}
\label{m2}
R(r)=r(1-\frac{r_{-}}{r})^{ a^2/(1+a^2)},~~\sigma(r)=1,~~
N(r)=(1-\frac{r_{+}}{r})(1-\frac{r_{-}}{r})^{(1-a^2)/(1+a^2)},
\end{eqnarray}
where $a$ is the dilaton coupling constant
($a=0$ corresponding to the Reissner-Nordstr\"om case),
$r_+,~r_{-}$ being  positive constant related to the configuration's mass and electric charge
(with $r_h=r_+>r_-$).
The periodicity $\beta$ of the Euclidean time coordinate, which
enters (\ref{action-tot}), depends on the 
parameters of the solution ($e.g.$ $\beta=4\pi r_h$ for a Schwarzschild geometry).

Another interesting background is provided by 
the euclideanized "hairy" black hole solutions in Einstein-Yang-Mills (EYM) theory
found in \cite{89}.
These purely magnetic solutions 
solve the EYM field equations also for an 
Euclidean signature \cite{Moss:1992ne}. 
In this case it is convenient to take $R(r)=r$ while no closed form expression
for the metric functions $N(r)$ and $\sigma(r)$ exists in the literature.
The numerical results we found give evidence that this background
support a second YM field.
This self-dual field  has a vanishing energy-momentum tensor
and does not curve the geometry, nor interact with the first, purely magnetic nonself-dual
gauge field.
  
In all cases, 
the gauge functions $w(r)$ and $u(r)$  
interpolate monotonically 
between the corresponding values at $r=r_h$ and the asymptotic values at infinity, 
without presenting any local extrema (this behaviour can be proven  analitycally from (\ref{SD-eq})).
For small enough values of $r_h$, the solutions 
look very similar to the flat space self-dual YM configuration.
These solutions get deformed increasing the value of $r_h$, while the value 
of the magnetic potential $w$ at $r=r_h$ steadly decreases.
As $r_h$ approaches some maximal value implied by (\ref{cond}), we find $w_h\to 0$ and the 
solution approaches
the limiting abelian configuration
\begin{eqnarray}
\label{ab}
w(r)=0,~~u(r)=\Phi+\int \frac{\sigma(r)}{R^2(r)}~dr.
\end{eqnarray}
For the case of an extremal background with $N(r)\sim O(r-r_h)^2$ as $r \to r_h$ 
($\sigma(r)$ and $R(r)$
being nonzero and finite in the same limit), we find
the trivial solution $w=1, u=\Phi$ only.

In Figures 1 and 2 we plotted several relavant parameters of the 
YM instanton solutions
as a function of $r_h$, for the case of Schwarzschild 
and  Reissner-Nordstr\"om backgrounds.
A typical self-dual YM solution in a 
euclideanized one-node EYM "hairy" black hole is plotted in Figure 3.
The event horizon radius in this case is $r_h=1$, while $w_h\simeq 0.9118$,
the periodicity of Euclidean time-coordinate $\tau$ being $\beta\simeq 19.644$.
These plots retain the generic features of the picture we found in other cases.

{\bf Self dual instantons in a topologically trivial background--}
The EYM "hairy" black hole solutions discussed above survive
in the limit $r_h \to 0$, when the particle-like 
Bartnik-McKinnon configurations,
indexed by the node number $k$ of the "primary" gauge field \cite{Bartnik:1988am} are approached.
This suggests the existence of self dual instantons for a 
topologically $R^4$ background,
the Killing vector 
$\partial/\partial \tau$ presenting in this case no fixed points sets 
($i.e.$ $g_{\tau \tau}>0$ for any $r$ and an arbitrary
periodicity $\beta$).
This type of backgrounds usually corresponds to analytical
continuations of Lorentzian globally regular, particle-like solutions.
A convenient choice here is $R(r)=r$
and we have $0\leq r <\infty$, while $N(0)=1$, $\sigma(0)=\sigma_0 \neq 0$.

The approximate expression of the YM instanton solutions as $r \to 0$ is
\begin{eqnarray}
\label{origin}
w(r)=1-br^2+O(r^4),~~
u(r)= 2b\sigma_0r+O(r^2),
\end{eqnarray}
(with $b>0$),
the asymptotic form (\ref{inf}) being valid in this case, too.

We solved the self-duality equations for the case of a 
a euclideanized Bartnik-McKinnon 
background with $k=1,2,3$ nodes. 
The profiles of the gauge functions $u$ and $w$ for the one-node 
Bartnik-McKinnon 
background are plotted in Figure 3.
The parameter $b$ which enters the expansion at the origin (\ref{origin})
is $b\simeq 0.003504$.

We found also solutions with similar 
properties for a gravitating skyrmion background \cite{Bizon:1992gb}.
Nontrivial solutions of the eqs. (\ref{SD-eq}) are likely to exist for other
topologically $R^4$ backgrounds.
The action $S$ of these solutions is still given by 
(\ref{action-tot}).
However, different from the bolt case, 
$S$ is independent on the background metric, since the 
parameter $\beta$ is arbitrary for a $R^4$ topology
(although the profiles of $u$ and $w$ depend on the
 details of the geometry).

{\bf Further remarks--}
In this letter we have presented numerical arguments for the existence of  
self-dual instantons for a number of spherically symmetric
backgrounds with Euclidean signature.
Hopefully, these numerical results will be of help in constructing
the solutions analytically.

Based on our results, we conjecture the existence of 
similar solutions for any spherically symmetric background
with no dependence on the Euclidean time function, which
approaches at infinity the euclideanized Minkowski background.

The existence of these solutions raise several interesting questions, since
in principle the Euclidean action of any spherically symmetric 
configuration can be adjusted to arbitrary values by including 
self-dual nonabelian fields in the action principle.

A discussion of possible generalizations of this work should start with
axially symmetric generalizations,  by including 
a winding number $n$ in the YM ansatz \cite{Rebbi:1980yi}.
Our preliminary numerical results indicate the existence
 of an axially symmetric YM self-dual solution in a Schwarzschild background,
 whose action is $n-$times the action of single self-dual (anti)instanton.
Such configurations are likely to exist for other background choice as well.
Also, it would be interesting to look for self-dual 
instantons in an axially symmetric background
($e.g.$ Kerr-Newman instanton).

Spherically symmetric YM  self-dual solutions
exist also for a different asymptotic structure of the background metric. 
For example, we found solutions with many similar properties for 
an Euclidean anti-de Sitter  background with $N(r)=1+r^2/\ell^2$, $\sigma(r)=1$
and $R(r)=r$ in (\ref{metric1}).

Further details on these solutions including an existence proof 
for a Schwarzschild background will be presented elsewhere.
\\
\\
{\bf Acknowledgement}
\\
We would like to thank D. H. Tchrakian for many useful discussions
and comments.
 This work is carried out
in the framework of Enterprise--Ireland Basic Science Research Project
SC/2003/390 of Enterprise-Ireland.  YB is grateful to the
Belgian FNRS for financial support.
\newpage

\end{document}